\documentstyle{article}
\begin{document}

\null
\vspace{2cm}
\begin{center}
\begin{Large}
\begin{bf}
Gravitational Microlensing by the MACHOs of the LMC\\
\end{bf}
\bigskip
\bigskip
\bigskip
\bigskip

Xiang-Ping Wu$^{1,2}$ \\
\end{Large}
\begin{large}

\bigskip
\bigskip

$^{1}$DAEC, Observatoire de Paris-Meudon, 92195 Meudon Principal Cedex,
         France\\
\bigskip
and\\
\bigskip
$^{2}$Beijing Astronomical Observatory, Chinese Academy of Sciences,\\
Beijing 100080, People's Republic of  China\\

\end{large}

\bigskip
\bigskip
\bigskip
\bigskip
\bigskip
\bigskip

\begin{large}
Submitted to\\

\bigskip
\bigskip

{\bf The Astrophysical Journal}\\

\bigskip
\bigskip
\bigskip
\bigskip

(Received \quad\quad\qquad\qquad )
\end{large}

\end{center}

\bigskip
\bigskip
\bigskip
\bigskip
\bigskip
{\noindent}-------------------------------------------------------------------------------------\\
{\noindent}Postal address: Xiang-Ping Wu$^{1}$ (before February 1994);
 Xiang-Ping Wu$^2$ (From February 1994).\\
{\noindent}E-mail: wxp@mesiob.obspm.circe.fr

\newpage

\begin{center}
\begin{Large}
\begin{bf}
Gravitational Microlensing by the MACHOs of the LMC\\
\end{bf}
\end{Large}

\bigskip
\bigskip

\begin{Large}

Xiang-Ping Wu \\

\bigskip
\begin{large}
DAEC, Observatoire de Paris-Meudon, 92195 Meudon Principal Cedex,
         France\\
and\\
Beijing Astronomical Observatory, Chinese Academy of Sciences,\\
Beijing 100080, People's Republic of  China\\
\end{large}
\bigskip
\bigskip
\bigskip
\bigskip
\bigskip
\bigskip
\bigskip

{\bf ABSTRACT}\\
\end{Large}
\end{center}

The expected microlensing events of the LMC by the MACHOs of the LMC
itself are calculated and compared with analogue events by objects
in the Galactic halo. The LMC matter distribution is modelled by
a spherical halo and an exponential disk while a face-on exponential
disk is used for the stellar distribution of the LMC.
Among the microlensing events discovered by the MACHOs and EROS projects,
a fraction of $22\%$ could be caused by the lenses near the center of
the LMC or $13\%$ from lenses at $5^o$ from the LMC center.
Therefore, any statistical study of these microlensing events must
take the LMC lenses into account.\\

\bigskip
\bigskip
\bigskip
\bigskip
{\noindent}-----------------------------------------------------------------------\\
{\noindent}{\bf\it Subject headings:} ~~gravitational lensing -- galaxies:
Magellanic
                              Clouds\\

\newpage
\begin{large}

\begin{center}
\begin{Large}
{\bf 1. ~INTRODUCTION}\\
\end{Large}
\end{center}

The recent detection of microlensing of stars in the Large Magellanic Cloud
(LMC) (Alcock et al., 1993; Aubourg et al., 1993) and in the direction of the
Galactic bulge (Udalski et al., 1993) opens a new way to search for
unseen matter in the halo and the disk of our Galaxy. Indeed,
these discoveries show that it has become possible nowadays to use
microlensing for the determination of mass range of compact objects
(MACHOs) in the Galaxy.  The above three experiments are still underway and
a statistically significant sample of microlensing events in both the LMC and
the Galactic bulge will hopefully be available in a few years.\\

It was realised a long time ago that nonstellar objects within
the Galaxy can act as gravitational lenses (Liebes, 1964).
However, the observational detection of these nonstellar population
(MACHOs) in the massive halo of our Galaxy using the gravitational lensing
effect had remained impossible until recent years when
Paczy\'nski (1986) proposed to monitor the brightness of several million
stars in the LMC, as the gravitational effects of the MACHOs can temporarily
magnify the light from  stars of LMC.  Griest (1991)
further developed this idea by adopting a more realistic mass distribution
model  for the MACHOs of the Galactic halo.
These two recent important papers have led people to believe that the three
microlensing events detected in the LMC are due to the MACHOs in the halo
of our Galaxy.\\

However, the contribution of unseen matter in the LMC itself,
possibly in the form of the MACHOs,  to the microlensing events has not been
taken into account in the determination of mass for the MACHOs by the
MACHO Project and the EROS collaboration, mainly because this effect is
believed
to be relatively small and can then  be negligible.
Gould (1993) has very recently considered the gravitational microlensing effect
of MACHOs of the LMC halo and found  that the LMC halo may
lead to a variation of the optical depth to lensing as a function of
sky position. However, it is unlikely that one can actually see such a
variation
without considering the distribution of stellar population of the LMC.
Unlike the one by the Galactic halo, the optical depth to microlensing
by the LMC halo/disk increases dramatically along the line of sight passing
through the LMC disk. Therefore, the expected microlensing events by the LMC
cannot be estimated simply by multiplying the optical depth by the total
number of stars of the LMC as for the case of the Galactic halo
(Paczy\'nski, 1986).  In fact, the optical depth to lensing by the LMC
halo/disk cannot provide valuable information about the fraction of
microlensing events produced by the LMC in the experiments, and
it has remained very unclear today how large the contribution
of the LMC halo and/or disk is to the detected events in the LMC.\\

In light of the upcoming statistical study of microlensing and its
significance for  searches of dark matter in modern astrophysics, this
paper presents a calculations of
the fraction of the expected microlensing events from the MACHOs
of the halo/disk of the LMC by taking into account  the distribution of
stellar population of the LMC modelled by a face-on exponential
disk. A direct comparison of microlensing contributions
from the halo of our Galaxy and from the halo/disk of the LMC will be given.\\

\newpage

\begin{center}
\begin{Large}
{\bf 2. ~OPTICAL DEPTH TO LENSING}\\
\end{Large}
\end{center}

\begin{center}
{\it 2.1~ Halo Model}\\
\end{center}

A rotational analysis of various populations of the LMC
(HI, planetary nebulae, clusters, LPVs, etc.) (Rohlfs et al., 1984;
Meatheringham et al., 1988; Hughes et al., 1991; Schommer et al., 1992; etc.)
indicates that  the LMC rotation curve exhibits both solid body (out to
$\sim2^o$) and flat rotation (to $8^o$--$15^o$) components. To fit this
rotation curve and also to simplify the calculation,  a spherical isothermal
halo with a definite core radius ($r_c$) is assumed for the LMC:
\begin{equation}
\rho(r)=\frac{\rho_0}{1+(r/r_c)^2}.
\end{equation}
Because such a halo extends to infinity and the mass diverges, it is
necessary to truncate the halo at some radius $R$ so that the total mass
within $R$ is
\begin{equation}
M(R)=4\pi\rho_0r_c^3\left(\frac{R}{r_c}-\arctan\frac{R}{r_c}\right),
\end{equation}
and the rotation velocity is
\begin{equation}
v(R)=\left[\frac{4\pi G\rho_0r_c^3}{R}
\left(\frac{R}{r_c}-\arctan\frac{R}{r_c}\right)
\right]^{1/2}.
\end{equation}
\\

The lensing cross-section by a pointlike mass ($m$) is simply
\begin{equation}
\pi a_c^2=\frac{4\pi G m}{c^2}\frac{D_d D_{ds}}{D_s},
\end{equation}
where $a_c$ is the critical radius  (the ``Einstein radius"), and the
other parameters have their usual meanings. The optical depth to microlensing
is
the probability of finding a source within $a_c$, the microlensing tube.
Assuming that the MACHOs in the LMC follow the distribution of eq.(1), one
has the total optical depth
\begin{equation}
\tau=\tau_0\int_{x_h}^{x_s}\frac{x(x_s-x)dx}{x_s(1+x_c^2-2x\cos\beta+x^2)}.
\end{equation}
Here
\begin{equation}
\tau_0=\left(\frac{v}{c}\right)^2
\frac{1}{1-\frac{r_c}{R}\arctan\frac{R}{r_c}}
\end{equation}
and $\beta$ is the angle between the line of sight to
the center and to the source (star) of the LMC. All distances
are measured in unit of $D_c$, the distance to the center of the LMC, which
is $50.6$ kpc (McCall, 1993). The radius of the LMC ($x_R=R/D_c$)
and the distance to the source ($x_s=D_s/D_c$)
define the ``edge" of the LMC halo  at the direction of $\beta$,
\begin{equation}
x_h=\cos\beta-\sqrt{x_R^2-\sin^2\beta}.
\end{equation}
Integration of eq.(5) gives
\begin{equation}
\begin{tabular}{ll}
$\frac{\tau}{\tau_0}=$& $\frac{x_h}{x_s}-1+(\frac{1}{2}-\frac{\cos\beta}{x_s})
\ln \frac{1+x_c^2-2x_s\cos\beta+x_s^2}{1+x_c^2-2x_h\cos\beta+x_s^2}+$\\
    & $\frac{1+x_c^2+(x_s-2\cos\beta)\cos\beta}{x_s\sqrt{x_c^2+\sin^2\beta}}
\left(\arctan\frac{x_s-\cos\beta}{\sqrt{x_c^2+\sin^2\beta}}
     -\arctan\frac{x_h-\cos\beta}{\sqrt{x_c^2+\sin^2\beta}}\right)$.
\end{tabular}
\end{equation}
The variation of optical depth with the source positions,  $D_s$
and $\beta$, is shown in Figure 1  for a halo radius of
$15^o$ (see Schommer et al, 1992). The rotation
velocity has been taken to be $79$ km/s. The corresponding result for
the MACHOs of the Galactic halo (Paczy\'nski, 1986; Griest, 1991) is
also plotted for comparison.  \\

The contribution of the LMC halo to the optical depth depends
sharply on the positions of stars in the LMC: The stars within
a projected angular distance of $2^o$ from the center of the LMC may
be significantly affected by the microlensing of foreground MACHOs
of the LMC, especially if the stars are  behind the disk plane. A larger
probability (as high as $10^{-6}$) than the one ($5\times10^{-7}$) by
the MACHOs in the Galactic halo appears to be possible.
For stars in the disk plane but within $2^o$ of the center of the LMC,
the optical depth is $(1$ to $2)\times10^{-7}$,   making the total optical
depth be $(6$ to $7)\times10^{-7}$ in combination with the value for the
halo of our Galaxy.  Even for the stars with a projected angular distance
of $5^o$ from the center but behind the disk,
the optical depth could still reach $10^{-7}$.\\

\begin{center}
{\it 2.2~ Disk Model}\\
\end{center}

The kinematic analysis shows that the dynamics of the LMC are
dominated by a single rotating disk.  The foreground MACHOs in the disk
of the LMC may then affect the background stars. Recall that  the LMC is
very close to face-on despite  the uncertainties in  measurement of the tilt
($i\approx27^o$). Matter distribution of the LMC can be assumed to be an
isothermal self-gravitating disk having density
\begin{equation}
\rho=\rho_0 e^{-R/h} {\rm sech}^2(z/z_0)
\end{equation}
with an exponential disk of the scale length $h$ in the radial direction
and an isothermal sheet of the scale height $z_0$ in the $z-$ direction
(van der Kruit \& Searle, 1981). The central mass density $\rho_0$ can
be related to the maximum rotational velocity by (Freeman, 1970)
\begin{equation}
\rho_0=\frac{1}{4\pi G h z_0}\left(\frac{V_m}{0.62}\right)^2.
\end{equation}
\\

For simplicity, the inclination is taken to be $0^o$, i.e., a
face-on distribution of the disk matter.  Repeating the procedure of
the above calculation  by using the matter distribution
of eq.(9), one has the total optical depth to microlensing by the
LMC disk
\begin{equation}
\tau=\tau_0\int_0^{xs}\;\;\frac{x(x_s-x)}{x_s}
\;\;\;\;e^{-\frac{x\sin \beta}{(h/D_c)}}
\;\;\;\;{\rm sech}^2\left(\frac{1-x\cos\beta}{(z_0/D_c)}\right)\;\; dx,
\end{equation}
where
\begin{equation}
\tau_0=\left(\frac{V_m}{0.62c}\right)^2\left(\frac{D_c^2}{hz_0}\right).
\end{equation}
The distance parameters, $x$ and $x_s$, and the position of the
sources, $D_s$ and $\beta$,  have the same definitions as in the above
section. The numerical results are shown in  Figure 2 by choosing
$V_m=79$ km/s, $h=1.6$ kpc and $z_0=0.43$ kpc for the LMC (Freeman,
Illingworth \& Oemler, 1983; Hughers, Wood \& Reid, 1991). \\

Foreground Stars are nearly unaffected by the microlensing of
the MACHOs in the LMC because  there are very few intervening lensing
objects. However, larger optical depths to microlensing
for the stars behind disk are found in the central region of the LMC,
resulting mainly from the higher number density of the MACHOs in the disk.
A probability as high as $10^{-6}$ appears for stars at a distance
of $z=5$ kpc behind the central disk. In general, the stars within an angle
of $\sim3^o$ from the center of the LMC
and a distance of $z\sim4$ kpc behind the disk would have a probability
for microlensing $\sim10^{-7}$ comparable to the one for the MACHOs of
the halo of our Galaxy.     Except for the sharp decrease for the
foreground stars, the optical depth by the MACHOs in the disk
varies with $D_s$ in a manner similar to that calculated for
the MACHOs in the halo of the LMC.\\

\newpage

\begin{center}
\begin{Large}
{\bf 3. EXPECTED MICROLENSING EVENTS}\\
\end{Large}
\end{center}

Microlensing experiments of monitoring the brightness of stars of the LMC
provide the microlensed stars and their positions, which cannot be
related directly to the optical depth to lensing.
The totally expected microlensing events can be found by convolving
the optical depth with the stellar population of LMC and then
integrating along the line of sight. \\

A disk model similar to eq.(9) can be applied to the distribution
of stellar population of the LMC, except that $\rho_o$ cannot be determined by
the dynamical analysis  of eq.(10).  The density of these stellar
objects along the line of sight is shown in Figure 3. The expected events
per solid angle  ($d\Omega$) and per line of sight distance ($dD_s$)
in the direction of $\beta$ from the center of the LMC are then
\begin{equation}
\frac{dN}{d\Omega dD_s}=\tau(\beta,D_s)\rho(\beta,D_s)D_s^2
\end{equation}
The variations of the expected events with $D_s$ are shown
in Figure 4(a) and Figure 4(b) using the optical depth by the LMC halo and
the disk, respectively.  The result by the Galactic halo is also
displayed for comparison, which mainly follows the distribution of
stellar population due to its nearly constant probability ($\tau$)
at the LMC position (see Figure 1 and Figure 2).  Although both the LMC halo
and the disk exhibit large optical depths at large distances behind
the disk as shown in Figure 1 and Figure 2, the expected events are sharply
confined within the disk since very few stars can be found at that
large distance behind the disk.\\

The totally expected events in the LMC  per solid angle in the direction of
$\beta$ are plotted in Figure 5(a)
for the contributions of the Galactic halo, the LMC halo,   the LMC disk and
the LMC halo + disk, respectively. Moreover, Figure 5(b) illustrates
the ratio of the expected events by the LMC to the ones by the Galactic halo.
The decrease of the expected events with the angular distance $\beta$
that results from the distribution of  stellar population of the LMC is well
demonstrated by the curve of the Galactic halo because of its nearly unchanged
optical depth at the LMC position.  The steeper slopes of the curves for
the LMC halo, in
particular the LMC disk, are then due to the fewer populations of the MACHOs
with the increase of angular distance $\beta$. This leads to a decrease
function of the ratio of $dN_{\rm LMC}/dN_{\rm Galaxy}$ with $\beta$.  \\

Define a parameter that describes the fraction of contributions by the LMC
in the experiment of searching for the microlensing events of the LMC:
\begin{equation}
q=\frac{N_{\rm LMC}}{N_{\rm Galaxy}+N_{\rm LMC}}.
\end{equation}
Figure 6 shows the variation of $q$ against the angular distance $\beta$ by
summing the contribution of the LMC halo and that of the disk. It turns out
that  the LMC is capable of producing
$22\%$ to $13\%$ of the totally detected microlensing events
from  the center  to  the distance of $\beta=5^o$ of the LMC.  This implies
that
one out the five events to be observed within $\sim2^o$ of the LMC center may
be created by the MACHOs of the LMC itself. \\

\newpage
\begin{center}
\begin{Large}
{\bf 4. CONCLUSIONS}\\
\end{Large}
\end{center}

The MACHOs of the LMC itself are shown to be able to act as gravitational
microlenses for stars of the LMC, responsible for as high as $22\%$ of
the microlensing events detected and to be detected in the LMC.
Thus, it is necessary to take the LMC microlensing events into account in
the statistical study of the recent microlensing experiments (the MACHOs and
EROS projects) for monitoring the brightness of stars of the LMC.
The expected microlensing events vary with sky position, due to
both the distribution of stellar population  and the
contributions of the MACHOs of the halo/disk of the LMC.
It is then argued that the
recently reported microlensing events near the center of the LMC
should be considered seriously if they all have a Galactic halo origin.
The distinction of the events by the LMC itself from the observed
ones, though it is  difficult, would be of importance
for the mass determination of the MACHOs making up the
halo of our Galaxy.\\

The present conclusions are based on a spherical halo and an exponential
disk model for matter distribution and a face-on exponential
disk model for the distribution of stellar population of the LMC.
It would be necessary in the future to have a better estimate of
the expected events by  adopting  a more realistic model involving
the tilt of the disk of the  LMC although it is unlikely that the small
angle of inclination of the LMC disk would significantly change the
present conclusion. Nevertheless, the idea of involving the distribution
of stellar population of the ``target" galaxy  suggested in this
paper should be also applied to SMC, M31 and M33 for the estimates of
their own contributions to the microlensing events  to be detected
in the future experiments.\\

\end{large}

\bigskip
\bigskip
\bigskip
\bigskip


{\noindent}{\bf ACKNOWLEDGMENTS}\\

I have benefited from my discussion with Alain Bouquet, Yannick
Giraud-H\'eraud, Jean Kaplan and Charling Tao.
I am also grateful to
Marshall L. McCall and Laurent Nottale for reading the manuscript and for
helpful comments, and  CNRS and K.C.Wong Foundation for financial support.\\

\newpage
\begin{large}
\begin{center}
{\bf References}\\
\end{center}

\bigskip

{\noindent}Alcock, C. et al., 1933, Nat, 365, 621\\
{\noindent}Aubourg, E. et al., 1933, Nat, 623\\
{\noindent}Freeman, K.C., 1970, ApJ, 160, 811\\
{\noindent}Freeman, K.C., Illingworth, G., Oemler Jr., A., 1983, ApJ, 272,
488\\
{\noindent}Griest, K., 1991, ApJ, 366, 412\\
{\noindent}Gould, A., 1993, ApJ, 404, 451\\
{\noindent}Hughes, S.M.G., Wood, P.R., Reid, N., 1991, AJ, 101, 1304\\
{\noindent}Liebes, J{\small R}., S., 1964, Phys. Rev., 133, B835\\
{\noindent}McCall, M.L., 1993, ApJ, 417, L75\\
{\noindent}Meatheringham, S.J., Dopita, M.A., Ford, H.C.,  Webster, B.L.,
           1988, ApJ, 327, 651\\
{\noindent}Paczy\'nski, B., 1986, ApJ, 304, 1\\
{\noindent}Rohlfs, K., Kreitschmann, J., Siegman, B.C., Feitzinger, J.V.,
	   1984, A\&A, 137, 343\\
{\noindent}Schommer, R.A., Olszewski, E.W., Suntzeff, N.B., Harris, H. C.,
           1992, AJ, 103, 447\\
{\noindent}Udalski, A. et al, 1993, ACTA Astronomica, 43, 289\\
{\noindent}van der Kruit, P.C., Searle, L., 1981, A\&A, 95, 105\\

\end{large}

\newpage

\begin{large}

\begin{center}
{\bf Figure Captions}\\
\end{center}

{\noindent}{\it Figure 1} ~~Variation of the optical depth to microlensing
by the MACHOs in the LMC halo with the positions of sources of
the LMC: the distance $D_s$ and the angular distance $\beta$ from the
center of the LMC.
The core radius of the LMC is taken to be
$2^o$ and the whole size is truncated at the radius of $15^o$. The optical
depth
by MACHOs of the Galactic halo is shown by the  dotted line. \\

{\noindent}{\it Figure 2} ~~The same as in Figure 1 but for the disk model.
The dynamical parameters are: $V_m=79$ km/s; $h=1.6$ kpc and $z_0=0.43$ kpc.\\

{\noindent}{\it Figure 3} ~~Variation of stellar population of
the LMC along the line of sight.\\

{\noindent}{\it Figure 4} ~~Variations of the expected microlensing
events along the line of sight per solid angle and per line of sight
distance by the LMC halo (a) and by the LMC disk (b). The contribution
of the Galactic halo is plotted as dotted lines.\\

{\noindent}{\it Figure 5a} ~~The expected microlensing events per
solid angle in the direction of $\beta$. The contributions of
the Galactic halo, the LMC halo, the LMC disk and the LMC halo +
disk are shown. \\

{\noindent}{\it Figure 5b} ~~The ratios of the expected microlensing
events by the LMC (halo, disk and halo+disk) to the ones by the
halo of our Galaxy.\\

{\noindent}{\it Figure 6} ~~The fraction of the microlensing events
by the LMC (halo+disk) in the microlensing experiments:
$q=N_{\rm LMC}/(N_{\rm Galaxy}+N_{\rm LMC})$.

\end{large}
\end{document}